\documentstyle[11pt]{article}

\begin{document}

\title{Interacting Quintessence}
\author{Winfried Zimdahl\footnote{Fax: +49 7531 884266; Electronic address:
winfried.zimdahl@uni-konstanz.de}\\
Fachbereich Physik, Universit\"at Konstanz\\ PF M678,
D-78457 Konstanz, Germany\\
\ \\
Diego Pav\'{o}n\footnote{Electronic address:
diego@ulises.uab.es}\\
Departamento de F\'{\i}sica,
Universidad Aut\'{o}noma de Barcelona\\
08193 Bellaterra (Barcelona), Spain\\
and\\
Luis P Chimento\footnote{Electronic address:
chimento@df.uba.ar}\\
Departamento de F\'{\i}sica, Universidad de Buenos
Aires\\ 1428~Buenos Aires, Argentina}
\date{\today}
\maketitle

\begin{abstract}
We demonstrate that
a suitable coupling between a quintessence scalar field and a pressureless
cold dark matter (CDM) fluid  leads to a constant ratio of the energy  
densities of both components which is compatible with an accelerated  
expansion of the Universe.
\end{abstract}

\section{Introduction}
There is a growing consensus among astrophysicists that we live in an
accelerating Universe. On the one hand, high--redshift type Ia supernovae
(SNIa) are significantly fainter than expected in a decelerating model (such 
as the Einstein--De Sitter) \cite{supernovae}. Although the statistics is
still low and extinction by interstellar dust may partly account for their
low brightness and no conclusive model of evolution of SNIa and their
progenitors is still available, the acceleration scenario is gaining further
ground \cite{Riess01}. On the other hand, while measurements of the average  
mass density of
the Universe systematically fall below the critical density, about $0.3$ or
$0.4$ in critical units (see e.g. \cite{bahcall} and references therein),
the position of the first acoustic peak  in  the temperature anisotropy
power spectrum of the CMB strongly suggests that the total energy density
is critical or near critical \cite{boomerang}. Combining both results
one may rule out a flat matter--dominated universe (with $\Omega_M = 1$
and $\Omega_\Lambda = 0$) as well as an open universe with no cosmological
constant ($ \Omega_M = 0.3 $ and $\Omega_{\Lambda} = 0$) at high statistical
level \cite{combining}. More generally, one is led to conclude that
very likely (i) about two third of the energy of the Universe is ``dark"
(i.e., non-luminous and not subject to direct detection via dynamical methods),
and (ii) connected to this exotic and elusive energy must be a negative
pressure, able to violate the strong energy condition.

The immediate candidate for such exotic energy, a small cosmological
constant $\Lambda$, poses however an embarrassing question:
Why the energy density in cold dark matter (which in the absence
of interactions redshifts as $a^{-3}$, where $a(t)$ is the scale factor
of the homogeneous and isotropic metric) and the constant energy
associated to $\Lambda$ are of the same order precisely today?
For this to occur one must have fine--tuned initial conditions right
after the inflationary epoch. This constitutes the so--called
``coincidence problem" \cite{coincidence}. To overcome this hurdle
it was suggested that a nearly homogeneous but time depending scalar
field with negative pressure should replace $\Lambda$. This peculiar
field, widely known as ``quintessence", was independently introduced
by Ratra and Peebles \cite{RaPeeb} and Wetterich \cite{Wetterich} well before the 
supernovae results were even suspected.
Today a host of quintessence models are known both in the realm of
general relativity (see e.g., \cite{caldwell}, \cite{amendola}) and
in scalar--tensor theories \cite{narayan}.

The target of this letter is to clarify a specific aspect of the coincidence  
problem, namely to present an attractor type solution of the two-component  
dynamics which is characterized by a constant ratio of order unity of the  
energy densities of the CDM and quintessence components and at the same time  
admits an accelerated expansion of the Universe.
The basic ingredient of the corresponding model is to assume a coupling  
between CDM and the quintessence scalar field.
It is this assumption of an interacting quintessence component by which our  
analysis differs from most investigations in this field which assume an  
independent evolution of CDM and the scalar field.
``Coupled quintessence'' models have  been shown to be useful in handling  
the coincidence problem by Amendola et al. \cite{amendola}. While the models  
of these authors assumed a specific coupling from the outset,
our strategy here is different. We do not specify the coupling from the beginning. 
We {\it determine} its structure from the requirement that it shall admit a  
solution for the dynamics of the two-component system of CDM and quintessence  
with a constant ratio for the energy densities.
This strategy seems legitimate since there does not exist any microphysical  
hint on the possible nature of a coupling between CDM and quintessence.
It will provide us with a transparent phenomenological picture of the  
``final state'' of the cosmic dynamics (for a less bleak eschatological  
scenario see \cite{JBarr}),
leaving open, of course, the question of how this state is approached and  
whether or not our current Universe has already reached it.

\section{Scalar field plus cold dark matter}
Let us consider a two--component system with an energy momentum tensor
\begin{equation}
T_{ik} = \rho u_{i}u_{k}
+ p h_{ik}\ , \
\label{1}
\end{equation}
where $h _{ik}=g_{ik} + u_{i}u_{k}$ and
\begin{equation}
\rho = \rho_{S} + \rho_{M}\ ,
\quad
p = p_{S} + p_{M}\ .
\label{2}
\end{equation}
The subscript S refers to the scalar field component, the subscript M to the  
matter component (i.e. CDM).
The energy density and pressure of the scalar field are
\begin{equation}
\rho_{S}={1 \over 2}\dot{\phi}^{2}
+ V (\phi) \
\quad {\rm and} \quad
p_{S}={1 \over 2}\dot{\phi}^{2}
- V (\phi)\ ,
\label{3}
\end{equation}
respectively.
The  splitting (\ref{2}) implies that there is only one 4--velocity,
\[
u^{i} = u_{M}^{i} = u^{i}_{S}
= - {g ^{ij}\phi _{,j} \over \sqrt{-g^{ab}
\phi _{,a} \phi _{,b}}}
\ .
\]
($\phi _{,a}$ is assumed to be timelike.)
We postulate that the components do not evolve independently but
that there exists some interaction
between them, described by a source (loss) term $\delta $ in the energy balances 
\begin{equation}
\dot{\rho}_{M} + 3H \left(\rho_{M}+p_{M}\right) = \delta
\ ,
\label{4}
\end{equation}
and
\begin{equation}
\dot{\rho}_{S} + 3H \left(\rho_{S}+p_{S}\right) = -\delta
\ .
\label{5}
\end{equation}
The last equation is equivalent to
\begin{equation}
\dot{\phi}\left[\ddot {\phi} + 3H \dot{\phi} + V^{\prime}\right] = -\delta
\ .
\label{6}
\end{equation}
As already mentioned, we will not specify the interaction
from the outset but constrain $\delta $  by demanding that the solution to
(\ref{4}) and (\ref{5}) be compatible with a constant ratio between
the energy densities $\rho _{M}$ and $\rho _{S}$.
It is convenient to introduce the quantities
$\Pi_{M}$ and $\Pi_{S}$ by
\begin{equation}
\delta \equiv - 3H \Pi_{M}
\equiv 3H \Pi_{S}\ ,
\label{7}
\end{equation}
with the help of which we can write
\begin{equation}
\dot{\rho}_{M} + 3H \left(\rho_{M}+p_{M} + \Pi_{M}\right) = 0,
\
\label{8}
\end{equation}
and
\begin{equation}
\dot{\rho}_{S} + 3H \left(\rho_{S}+p_{S} + \Pi_{S}\right) = 0
\ .
\label{9}
\end{equation}
The rewriting of Eqs. (\ref{4}) and (\ref{5}) into Eqs. (\ref{8}) and  
(\ref{9}), respectively, makes the dynamic equations formally look as those  
for two dissipative fluids.
The fact that there is a coupling between them has been mapped onto
the relation $\Pi _{M}=-\Pi _{S}$ between the effective pressures $\Pi _{M}$  
and $\Pi _{S}$.
Some early models of power law inflation also share
this feature (see e.g., \cite{yokoyama}).
Below we shall map the interaction term $\delta $ onto a corresponding interaction potential.

\section{Attractor solution and cosmological dynamics}

Consider now the time evolution of the ratio $\rho _{M}/\rho _{S}$,
\begin{equation}
\left(\frac{\rho _{M}}{\rho _{S}} \right)^{\displaystyle \cdot}
= \frac{\rho _{M}}{\rho _{S}}
\left[\frac{\dot{\rho }_{M}}{\rho _{M}}
- \frac{\dot{\rho }_{S}}{\rho _{S}}\right]\ .
\label{10}
\end{equation}
By introducing the shorthands
\begin{equation}
\gamma _{M} \equiv  \frac{\rho _{M} + p _{M}}{\rho _{M}}
= 1 + \frac{p _{M}}{\rho _{M}}, \quad \mbox{and} \quad
\gamma _{S} \equiv  \frac{\rho _{S} + p _{S}}{\rho _{S}}
= \frac{\dot{\phi }^{2}}{\rho _{S}} \  ,
\label{11}
\end{equation}
we obtain
\begin{equation}
\left(\frac{\rho _{M}}{\rho _{S}} \right)^{\displaystyle \cdot}
= - 3H \frac{\rho _{M}}{\rho _{S}}
\left[\gamma _{M} - \gamma _{S}
+ \frac{\rho }{\rho _{M}\rho _{S}}\Pi _{M}\right]\ .
\label{12}
\end{equation}
Obviously, there exists a stationary solution
$\left(\rho _{M}/\rho _{S} \right)^{\displaystyle \cdot} = 0$ for
\begin{equation}
\Pi _{M}=-\Pi _{S}=\frac{\rho _{M}\rho _{S}}{\rho _{M} + \rho _{S} }
\left(\gamma _{S} - \gamma _{M}\right)\ .
\label{13}
\end{equation}
Since the CDM behaves as dust, i.e. $p _{M}\ll \rho _{M}$, we find
\begin{equation}
\Pi _{M} \approx- \left[1 - \frac{\dot{\phi }^{2}}
{\frac{1}{2}\dot{\phi }^{2} + V}\right]\frac{\rho _{S}\rho _{M}}{\rho }\ ,
\label{14}
\end{equation}
or, by virtue of $\frac{1}{2}\dot{\phi }^{2} - V = p _{S} \approx p$,
\begin{equation}
\Pi _{M} \approx \frac{\frac{1}{2}\dot{\phi }^{2} - V}
{\frac{1}{2}\dot{\phi }^{2} + V}\frac{\rho _{S}}{\rho }\rho _{M}
= \frac{p}{\rho }\rho _{M}\ .
\label{15}
\end{equation}
The coupling term corresponding to this is
\begin{equation}
\delta = 3H\left[1 - \frac{\dot{\phi }^{2}}
{\frac{1}{2}\dot{\phi }^{2} + V}\right]\frac{\rho _{S}\rho _{M}}
{\rho} \   ,
\label{16}
\end{equation}
or, equivalently,
\begin{equation}
\delta = - 3H \frac{p _{S}}{\rho }\rho _{M}
= -3 H \left(\gamma _{S} - 1\right)
\frac{\rho _{S}\rho _{M}}{\rho _{S} + \rho _{M} }.
\label{17}
\end{equation}
Introducing the notation $r \equiv   \rho _{M}/\rho _{S}={\rm const}$
we may further write
\begin{equation}
\delta = - 3H \left(\gamma _{S} - 1\right)\frac{\rho _{M}}{r+1},
\quad \mbox{or} \quad
\delta = - 3H \left(\gamma _{S} - 1\right)\frac{r}{r+1}\rho _{S}  \  .
\label{18}
\end{equation}
Invoking the Friedmann equation valid for universes with
spatially flat sections,
\begin{equation}
3H^{2}= 8\pi G \left[\rho _{S} + \rho _{M}\right] \  ,
\label{19}
\end{equation}
we have
$3H = \sqrt{24\pi G \rho }$, and, consequently,
\begin{equation}
\delta = - \sqrt{24\pi G }
\left(\gamma _{S}-1 \right)
\frac{\rho _{S}\rho _{M}}{\sqrt{\rho _{S}+\rho _{M}} }\ .
\label{20}
\end{equation}
With (\ref{18}), in a spatially flat universe equivalent to (\ref{20}), we  
have identified the interaction between the pressureless fluid (CDM) and the  
scalar field (quintessence) that guarantees a constant ratio $r$ of the  
energy densities of both components.

To study the stability of this stationary solution against small
perturbations we introduce the ansatz
\[
\frac{\rho _{M}}{\rho _{S}} = \left(\frac{\rho _{M}}{\rho _{S}} \right)_{st} 
+ \epsilon \
\]
into (\ref{12}) -the subscript {\small st} is for ``stationary".
The result is 
\begin{eqnarray}
\dot{\epsilon } &=& 3H \left[\left(\frac{\rho _{M}}{\rho _{S}} \right)_{st}  
+ \epsilon \right]\left[\frac{p _{S}}{\rho _{S}}
- \frac{\rho }{\rho _{S}}\frac{\Pi _{M}}{\rho _{M}}\right]\nonumber\\
&=& 3H \left[\left(\frac{\rho _{M}}{\rho _{S}} \right)_{st} + \epsilon  
\right]\left[\frac{p _{S}}{\rho _{S}}
- \left(1 + \left(\frac{\rho _{M}}{\rho _{S}} \right)_{st} + \epsilon
 \right)\frac{\Pi _{M}}{\rho _{M}}\right]
\ .
\label{20a}
\end{eqnarray}
The behavior of the perturbed solution depends on the ratio
$\Pi _{M}/ \rho _{M}$.
For the stationary solution itself we may read off $\Pi _{M}$ from (\ref{7})  
and (\ref{18}).
However, for deviations from stationarity an additional assumption is necessary. 
At first sight the most obvious choice seems to be
$|\Pi _{M}| \propto \rho _{M}$ also in the vicinity of the stationary  
solution. As to be seen from (\ref{7}), the coupling term becomes asymmetric  
with respect to $\rho _{M}$ and
$\rho _{S}$  under such conditions.
It will turn out that a more appropriate choice is the assumption
$\Pi _{M} = - c \rho $, where $c$ is a constant $c>0$.
This type of interaction is symmetric in $\rho _{M}$ and
$\rho _{S}$.
Up to first order in $\epsilon $ we find in such a case,
\begin{equation}
\dot{\epsilon } = 3Hc  \frac{r ^{2}-1}{r}\epsilon\ .
\label{20b}
\end{equation}
This implies that the stationary solution is stable for $r<1$, which is  
clearly compatible  with the presently favored observational data
$\rho _{M} \approx 0.3$ and $\rho _{S} \approx 0.7$.
Consistency with $\Pi _{M}$ from (\ref{7}) and (\ref{18}) fixes c:
\begin{equation}
c = r \frac{1-\gamma _{S}}{(1+r)^2} \ .
\label{20c}
\end{equation}
The positivity of  $c$ is guaranteed for $\gamma _{S}<1$.

With $p \approx p_S$ today, the stability condition corresponds to
[cf. Eq. (\ref{20a})]
\begin{equation}
\frac{p}{\rho } - \frac{\Pi _{M}}{\rho _{M}} \leq  0 \ .
\label{21}
\end{equation}

\noindent
Since we seek accelerated expansion, the total pressure $p \approx p _{S}$  
must be negative, i.e., the potential term must dominate the kinetic term,  
equivalent to $\gamma _{S} < 1$. 
It is remarkable that according to (\ref{20c}) this coincides with  
the condition for $c$ to be positive. 
From (\ref{7}) and (\ref{18}) we find that
a value $\gamma _{S}<1$ implies $\Pi_M < 0 $ and $\delta > 0$.
There is a tranfer of energy from the scalar field to the matter, 
which reminds of decaying vacuum energy approaches for the dynamics 
of the early universe (see, e.g., \cite{Freese}). 
The stationary epoch
$\Pi _{M}/\rho _{M}=p/\rho $ has to be approached in such a way that
\begin{equation}
\frac{|\Pi _{M}|}{\rho _{M}} \leq \frac{|p|}{\rho }\ .
\label{22}
\end{equation}
Since $|\Pi _{M}|$ is proportional to $\delta $, this means, the
interaction may be small as long as the system is still far from the
attractor solution.

It is expedient to emphasize that the apparently subtle point to assume 
$|\Pi _{M}| \propto \rho $ instead of
$|\Pi _{M}| \propto \rho _{M}$ is essential for the stability properties of  
the stationary solution.
Namely, similar considerations as those leading to (\ref{20b}) show, that  
there does not exist a stable solution with accelerated expansion for
$|\Pi _{M}| \propto \rho _{M}$.
Therefore, a dependence $|\Pi _{M}| \propto \rho $ is mandatory for a  
physically sensible solution.
This represents a restriction on the type of interaction that produces a  
stationary ratio $\rho _{M}/\rho _{S}$ .
While for the stationary solution itself $\Pi _M \propto  \rho _M$
and $\Pi _M \propto  \rho $ are not really different since
$\rho _M \propto  \rho $, the difference becomes crucial if one perturbs the  
solution. 

Note that the stability is connected to the presence of
an effective  dissipative
stress in the matter fluid. This parallels the result
that the scalar field needs the assistence of a dissipative fluid
stress for the coincidence problem to find solution in spatially flat
accelerating Friedmann--Robertson--Walker models \cite{chimento}.

Given the interaction term  (\ref{18}),  we may find  the dependence of
$\rho _{M}$
and $\rho _{S}$ on the scale factor. Because of $p_M \approx 0$, 
Eq. (\ref{4}) 
with (\ref{18}) yields
\begin{equation}
\dot{\rho }_{M} + 3 H \rho _{M} = - 3H \left(\gamma _{S} -1\right)
\frac{\rho _{M}}{r+1}\ ,
\label{23}
\end{equation}
while (\ref{5}) with (\ref{18}) results in
\begin{equation}
\dot{\rho }_{S} + 3 H \gamma _{S}\rho _{S} = 3H \left(\gamma _{S} -1\right)
\frac{r}{r+1}\rho _{S}\ .
\label{24}
\end{equation}
Assuming $\gamma_{S}$, which is in the range
$0 \leq \gamma _{S} \leq 2$, to be (at least piecewise) constant, we obtain
\begin{equation}
\rho _{S} \propto a ^{-\nu }\ , \quad
\rho _{M} \propto a ^{-\nu }\ ,
\quad \nu = 3 \frac{\gamma _{S}+r}{r + 1}\ .
\label{25}
\end{equation}
Both energy densities happen to redshift at the same rate because we have
chosen $\delta $ to correspond to the stationary state.
With the relationship $\rho \propto a ^{- \nu }$ we can solve
the Friedmann equation (\ref{19})  to find
\begin{equation}
a \propto t ^{2/\nu }
\quad\Rightarrow\quad q \equiv  -\frac{\ddot{a}}{aH ^{2}}
= - \left(1 - \frac{\nu }{2} \right)\ .
\label{26}
\end{equation}
The total energy density redshifts as $\rho \propto t ^{-2}$, independently
of $\gamma _{S}$  and $r$. Power law accelerated expansion will occur for
$\nu < 2$, equivalent to
\begin{equation}
 r + 3 \gamma _{S} < 2 \  .
\label{27}
\end{equation}

\noindent
Together with the above derived stability condition $r < 1$ this  
amounts to
$\gamma _{S}< 1/3$ or $p _{S}/\rho _{S}< -2/3$ for accelerated expansion.

Defining
\begin{equation}
\Omega _{M}\equiv  \frac{8\pi G}{3H ^{2}}\rho _{M}\ ,\quad {\rm and}\quad
\Omega _{S}\equiv  \frac{8\pi G}{3H ^{2}}\rho _{S}\ ,
\label{27a}
\end{equation}
we have
\begin{equation}
\Omega _{M} = \frac{r}{r+1} \ , \quad {\rm and} \quad
\Omega _{S} = \frac{1}{r+1}\ ,
\label{27b}
\end{equation}
respectively, and also
\begin{equation}
\Omega _{S} = \frac{8\pi G}{3}\frac{\nu ^{2}}{4}\rho _{S}t ^{2}\ .
\label{27c}
\end{equation}
For $\rho _{S}$  we find
\begin{equation}
\rho _{S} = \frac{1}{6\pi G }
\frac{1+r}{\left(\gamma _{S}+ r \right)^{2}}\ \frac{1}{t ^{2}}\ .
\label{28}
\end{equation}
Combination with (\ref{11}) yields
\begin{equation}
\dot{\phi } = \sqrt{\frac{\gamma _{S}\left(1+r \right)}{6\pi G }}
\frac{1}{\left(\gamma _{S}+r \right)}\ \frac{1}{t}
\ ,
\label{29}
\end{equation}
i.e., $\phi $ evolves logarithmically with time.
Furthermore, with the help of (\ref{3}) and (\ref{11}) it follows that
\begin{equation}
\rho _{S} = \frac{2V(\phi)}{2-\gamma_{S}} = \frac{\dot{\phi }^{2}}
{\gamma _{S}}\ ,
\label{30}
\end{equation}
which together with (\ref{28}) and (\ref{29}) leads to
\begin{equation}
V(\phi) = \frac{1}{6\pi G }\left(1 - \frac{\gamma _{S}}{2}\right)
\frac{1+r}{\left(\gamma _{S}+r \right)^{2}}\ \frac{1}{t ^{2}}\ .
\label{31}
\end{equation}
Since
\begin{equation}
V ^{\prime }(\phi)\dot{\phi } = \dot{V}(\phi) = - 2\frac{V}{t}\ ,
\label{32}
\end{equation}
by virtue of (\ref{29}) we obtain
\begin{equation}
V ^{\prime }(\phi) = - \lambda V(\phi) \ ,
\label{33}
\end{equation}
where
\begin{equation}
\lambda =
\sqrt{\frac{24\pi G}{\gamma _{S}\left(1+r \right)}}
\left(\gamma _{S} + r\right)
\label{33a}
\end{equation}
and, consequently,
\begin{equation}
V(\phi) = V _{0}\exp{\left[- \lambda  \left(\phi - \phi _{0} \right)\right]}\ .
\label{34}
\end{equation}
By similar steps one shows that the interaction term $\delta $ in 
Eq. (\ref{6}), given by the second expression in (\ref{18}),  may be mapped onto an interaction potential $V _{int}$: 
\begin{equation}
\frac{\delta }{\dot{\phi }}\equiv  V _{int}^{\prime } 
\quad\Rightarrow\quad 
V _{int} = - \frac{2r}{\gamma _{S}+r}
\frac{1- \gamma _{S}}{2 - \gamma _{S}}V(\phi) 
\ .
\label{34a}
\end{equation}
Introducing an effective potential 
\begin{equation}
V _{eff} \equiv  V(\phi) + V _{int}\ ,
\label{34b}
\end{equation}
the equation of motion for the $\phi $ field becomes 
\begin{equation}
\ddot {\phi} + 3H \dot{\phi} + V _{eff}^{\prime}= 0
\ .
\label{34c}
\end{equation}
It is rather reassuring
(although not unexpected, cf.(\cite{LuMa}) to find a potential (\ref{34b})  
with (\ref{34}) and (\ref{34a}),  substantially
backed by some field theories. It appears for instance in $N = 2$
supergravity \cite{reassuring1}. Likewise, linear combinations of
exponential potentials naturally arise in theories undergoing dimensional
compactification to an effective 4-dimensional theory; it is reasonable
to expect that one of them will eventually dominate \cite{reassuring2}.

With the help of (\ref{27}) the  condition for accelerated expansion becomes 
\begin{equation}
\lambda ^{2} < 24\pi G
\frac{\left(1 - \gamma _{S}\right)^{2}}{\left(1+r \right)\gamma _{S}}\ .
\label{35}
\end{equation}
This is similar but not identical to conditions which have been obtained for  
corresponding solutions in the non-interacting case  
\cite{Wetterich,CoLiWands,Ferreira} or for different types of coupling
\cite{amendola,yokoyama,CoLiWands,BillCol}.
These authors started with an exponential potential in which $\lambda $ is a  
free parameter initially. Then they investigated the parameter range for  
which there exists an attractor solution which is also inflationary.
Our strategy is different insofar, as we have first constructed a solution  
with the required properties and then read off the corresponding parameter  
combination. 

Notice also that the way the attractor is approached remains open (only
that in order to guarantee stability the approach, according to
(\ref{22}), has to proceed from a smaller coupling than given
by the stationary solution itself).

\section{Discussion}

We proposed a coupling $\delta$ (given by (\ref{17}), (\ref{18}),  
or (\ref{34a}) with (\ref{34})) between  a quintessence
scalar field and a CDM fluid that leads to a stable, constant ratio for the
energy densities of both components, compatible with a power law accelerated 
cosmic expansion.
This interacting quintessence approach indicates a phenomenological solution of  
the coincidence problem that afflicts many
attempts to cope with late acceleration (especially those based in a
cosmological constant). Unlike other approaches the potential 
is not an input but derived from the coupling.
It remains to be seen
to what extent this potential is consistent
with  measurements of the supernovae distances \cite{distances} once
the SNAP satellite comes up with enough SNIa statistics \cite{snap}.
Alternative and possibly earlier available tests rely
on the Alcock-Paczy\'nsky test for quasar pairs [cf. Ref. \cite{dalal}].

While focusing on the stationary solution
straightforwardly provides us with an expression for the interaction which  
realizes a corresponding state, we mention again that this procedure leaves  
open how this interaction is exactly ``switched on'' in order to account for  
the necessary transition from the era of decelerated expansion to that of  
accelerated expansion.
The coupling should be ineffective until the condensation of protogalaxies  
has entered the non-linear regime.
In a sense, this feature reminds of the ``exit problem'' of many  
inflationary models.
There are attempts to tackle this problem with the help of a specific  
coupling function between $\phi$ and CDM together with a separately  
postulated exponential potential \cite{amendola}. However, a really  
satisfactory solution is still missing.
What one would like to have is an interaction which is negligible in the  
matter dominated era and asymptotically approaches (\ref{17}) for large  
times.
We hope that our stationary solution  will give an indication for
a quintessence--CDM coupling that, aside from characterizing the stationary  
state of the late accelerated expansion, smoothly joins
the previous matter--dominated era of decelerated expansion when one goes  
backward in time.

\section*{Acknowledgments}
This work was supported by the Deutsche Forschungsgemeinschaft,
the Spanish Ministry of Science and Technology under grant
BFM 2000-0351-C03-01, the University of Buenos Aires under project
X223-2001-2002, and NATO.

\end{document}